\documentclass[prl,
aps,superscriptaddress,showpacs,preprintnumbers,nofootinbib,amsmath,amssymb,amsfonts,bm,marginnote,cite,graphicx,showlabels,
slashed,color, twocolumn]{revtex4}
\usepackage{graphicx,color}

\newcommand{\be}{\begin{equation}}
\newcommand{\ee}{\end{equation}}
\newcommand{\beq}{\begin{eqnarray}}
\newcommand{\eeq}{\end{eqnarray}}
\newcommand{\bea}{\begin{eqnarray}}
\newcommand{\eea}{\end{eqnarray}}
\newcommand{\beqn}{\begin{eqnarray}}
\newcommand{\eeqn}{\end{eqnarray}}

\def\tx{\tilde{x}}

\newcommand{\cE}{{\cal E}}

\newcommand{\rd}{\mathrm{d}}

\def\tp{\tilde p}

\def\cH{{\cal{H}}}

\def\tx{\tilde{x}}

\def\tp{\tilde{p}}

\begin{document}

\title{Non-Fermi Liquids, Strange Metals and Quasi-metaparticles}

\author{Edwin Barnes}
\affiliation{Department of Physics, Virginia Tech, Blacksburg, VA 24061, U.S.A.}

\author{J. J. Heremans}
\affiliation{Department of Physics, Virginia Tech, Blacksburg, VA 24061, U.S.A.}

\author{Djordje Minic}%\email{dminic@vt.edu}
\affiliation{Department of Physics, Virginia Tech, Blacksburg, VA 24061, U.S.A.}

\date{\today}

\begin{abstract}
We introduce the concept of {quasi-metaparticles} based on the theory of metaparticles, the zero modes of the metastring. We apply the concept of quasi-metaparticles to the problem of non-Fermi liquids and the properties of strange metals. In particular, we point out that the quasi-metaparticle Green's function interpolates between the canonical quasi-particle Green's function and the result found in the context of the SYK model, which presents an exactly solvable model without quasiparticles. The linear dependence of resistivity with temperature is reproduced in the SYK limit. Also, the Cooper mechanism is possible in the quasi-metaparticle case. Finally, the new parameter that characterizes quasi-metaparticles can be extracted from ARPES data. Thus, the quasi-metaparticle could be a useful new concept in the study of strange metals and high-temperature superconductivity.
\end{abstract}

\maketitle
%%%%%%%%%%%%%%%%%%%%%%%%%%%%%%%%%%%%%%%%%%%%%%%%%%%%%%
{\it Introduction:}
One of the central ideas of condensed matter physics is the concept of (Landau's) 
Fermi liquids.
This fundamental notion can be understood from the modern approach to quantum (effective) field theory in
the presence of the Fermi surface \cite{Polchinski:1992ed} (see also, \cite{abrikosov, anderson1}).
Tied to this central notion is the concept of quasi-particles.
However, this crucial feature of good metals is manifestly transcended in the context of strange metals \cite{varma, varmareview}. The strange metals do not support the concept of quasiparticles and some
of the well known models (such as the Sachdev-Ye-Kitaev, SYK model \cite{Sachdev:1992fk}) do not support quasiparticles at all.
The SYK model captures a
marginal Fermi liquid behavior \cite{varma, varmareview} of the Green's function.
The central phenomenological feature of strange metals is the linear dependence of the resistivity with temperature \cite{varmareview}.
Here we stress the point that existing models capture either Fermi liquid behavior or strange metal behavior (without quasiparticles), but not both. This motivates the question: Is there a theory in which both behaviors can arise as limiting cases?

In this letter we introduce the concept of {quasi-metaparticles} based on the theory of metaparticles, the zero modes
of the metastring \cite{Freidel:2017nhg, Freidel:2017wst, Freidel:2018apz, Freidel:2021wpl}. 
We apply the concept of quasi-metaparticles to the problem of non-Fermi liquids and
the properties of strange metals. In particular, we point out that the quasi-metaparticle propagator 
interpolates between the quasi-particle propagator and the propagator found in the context of
SYK model, which presents an exactly solvable model without quasiparticles. The linear dependence of
resistivity with temperature is reproduced in the SYK limit and  the Cooper mechanism is supported as well. Thus the quasi-metaparticle might be a 
useful new concept in the study of strange metals.

{\it The Metaparticle \cite{Freidel:2018apz}:} 
The concept of metaparticles was introduced in the context of string theory  \cite{Freidel:2017nhg, Freidel:2017wst} and
it was explicitly developed in \cite{Freidel:2018apz}. In particular, the new Born geometry \cite{Freidel:2013zga, Freidel:2014qna} was introduced 
to describe a target geometry of string theory upon which T-duality (target-space duality) acts as a linear symmetry 
\cite{Freidel:2015pka,Freidel:2017xsi}.
The main intuitive idea behind Born geometry is to preserve the duality between position 
and momentum space that we have in ordinary quantum mechanics
in the case of curved spacetimes.
In this context, locality itself can be observer (or probe) dependent, in other words, locality is relative   \cite{AmelinoCamelia:2011bm}.
Recently a fundamentally new model of space, called {modular space} \cite{Freidel:2016pls, Freidel:2017xsi}, was proposed as a template for a space incorporating these ideas organically into its fabric.
Modular spaces appear as a choice of polarization (a commutative subalgebra) of quantum Weyl algebras that do not have any classical analogue but possess a {covariant} built-in length scale. 
Modular spaces capture the essential non-locality of quantum theory that is consistent
with causality \cite{Freidel:2016pls}, as illustrated by modular variables \cite{aharonov},
such as the Aharonov-Bohm and Aharonov-Casher phases \cite{aharonov}.
The connection between modular spaces and string theory stems from the fundamental results that closed string theory target space is intrinsically non-commutative for the compactified modes \cite{Freidel:2017nhg, Freidel:2017wst}).
In \cite{Freidel:2018apz} the {metaparticle theory} was developed, which may be thought of as a particle model that retains the principle zero-mode structure of the string and thus propagates on a Born geometry. Alternatively, one can understand the metaparticle as a modular particle, the natural relativistic fundamental excitation supported by modular space and supported by the intrinsic quantum non-locality consistent with causality \cite{Freidel:2016pls}.

The metaparticle is defined  \cite{Freidel:2018apz} in a doubled target phase space of Lorentzian signature, whose coordinates we label $\{x^\mu,p_\mu,\tx_\mu,\tp^\mu\}$, with $\mu=0,1,...,d-1$. If we were to write down the ordinary particle action on this phase space, there would be a physical problem owing to the signature of the coordinate space. So, in addition to the usual Hamiltonian constraint, which now reads
%\beq
${\cal H}=\frac12(p^2+\tp^2+m^2)= \frac12\left(p_\mu h^{\mu\nu}p_\nu +\tp^\mu h_{\mu\nu}\tp^\nu + m^2\right)$,
%\eeq
we have a second constraint
%\beq
${\cal D}=p_\mu\tp^\mu-\mu$.
%\eeq
The worldline metaparticle action is given by
%\beq
$S=\int\rd \tau \left[p\cdot\dot x- {\tilde x}\cdot\dot{\tilde p}+\pi\alpha' p\cdot\dot{\tilde p}-e{\cal H}-\tilde e{\cal D}\right]$,
%\eeq
where $e$ and $\tilde{e}$ are the Lagrange multipliers, $\alpha'$ is the fundamental length squared and $\mu$ is a new dimensionful parameter.
In the analogous string theory, the ${\cal H}$ and ${\cal D}$ constraints are associated with world-sheet diffeomorphism invariance,
and setting them to zero on quantum states give the on-shell conditions for particle states
whose oscillator levels are associated with values of $m^2$ and $\mu$ \cite{Freidel:2017nhg, Freidel:2017wst}. Thus in the string theory analogue, the metaparticle theory corresponds (for specific values of $m^2,\mu$) to a formulation of the dynamics of particle states at a fixed level. 

{\it The Metaparticle Quantum Propagator \cite{Freidel:2018apz}:} 
To understand better what the metaparticle is, next we present an overview of the structure of the quantum propagator.  Since we have two constraints, in the $x,\tp$ polarization,
the quantum states will be further labelled by a pair of  evolution parameters, which we 
call $\ell,\tilde\ell$. The transition amplitude in the $x,\tp$ polarization has a canonical interpretation
\beq
K(x_f,\tilde p_f,\ell_f,\tilde{\ell}_f; x_i,\tilde p_i,\ell_i,\tilde{\ell}_i)
=\langle x_f,\tilde p_f;\ell_f,\tilde{\ell}_f| x_i,\tilde p_i;\ell_i,\tilde{\ell}_i\rangle \nonumber 
\\
=\langle x_f,\tilde p_f|e^{-i(\ell_f-\ell_i) \hat\cH-i (\tilde{\ell}_f-\tilde{\ell}_i)\hat{\cal D}}| x_i,\tilde p_i\rangle\label{MetaParticlePI}.
\eeq
Integrating over  $\ell\in (0,\infty)$ and $\tilde\ell\in (-\infty,\infty)$ then yields the metaparticle propagator  \cite{Freidel:2018apz}
\beqn\label{doubletramp}
G(x,\tilde p; 0,\tilde p_i)
\sim
\delta^{(d)}(\tilde p-\tilde p_{i})
\int \frac{d^dp}{(2\pi)^d}\frac{\delta(p\cdot\tilde p-\mu) e^{ip\cdot x} }{p^2+\tilde p^2+m^2-i\varepsilon}.
\eeqn
There are two differences compared to an ordinary relativistic particle propagator. One is the $\delta$-function of the ${\cal D}$ constraint, and the other is the presence of $\tilde p$ in the denominator. 
We note that the propagator is invariant under the change $\mu\to -\mu$, if we simultaneously change $\tp\to-\tp$. Consequently, we will without loss of generality assume that $\mu>0$.

In what follows we concentrate on the case when  $\tp$ is null.
(We will discuss the space-like and time-like cases at the end of this letter.)
In that case $\tp$  can be parametrized as  $\tp={\cal E}(1,\vec{n})$ with $\vec n^2=1$, and we find that the $\delta$-function fixes the light-cone momentum.  More precisely, we have
$p=p_-(1,\vec n)-\tfrac{\mu}{2\cal E} (1,-\vec n) +  (0,\vec{p}_\perp)$ for any $(p_-,\vec p_\perp)$ with $\vec{n}\cdot\vec p_\perp=0$. 
Ultimately,  we find  \cite{Freidel:2018apz}
\beqn\label{doubletrampLC}
G(x,{\cal E},\tilde{n})
\sim 
\int d^{d-2}p_\perp\int dp_-\frac{e^{i\left(-\tfrac{\mu}{\cal E} x^++ p_-x^-  + \vec p_\perp\cdot\vec x_\perp\right)}}{-2\mu p_-+{\cal E}{\vec p}_\perp^{\,2}%+m^2+\frac{\mu^2}{{\cal E}^2}
-i\varepsilon} 
\eeqn
which contains a single non-relativistic pole with an effective mass $\mu/{\cal E}$. 
Note that when $\mu$ is zero, it is  possible to consider only external states with  $\tp=0$. In this case, the $\delta$-function has no restriction on the particle momenta  and the pole is the same as in the relativistic particle. 
This shows that the usual relativistic particle can be viewed as a metaparticle which is ``massless''  for the $\cal D$ constraint and which possesses vanishing external dual momentum. 
This sector corresponds to a consistent truncation of the theory since the vanishing of $\tp$ is consistent with momentum conservation and is preserved by the interactions. In the analogue string theory, this is the consistent truncation  of the spectrum that occurs in the decompactification limit.

{\it Quasi-Metaparticles, Non-Fermi Liquids and Strange Metals:}
We can now state our main point regarding the non-Fermi-liquid form of the metaparticle propagator,
in the context when there is a Fermi surface present and when all energies are scaled towards the Fermi surface 
(which might be very irregular) \cite{Polchinski:1992ed}.
We will only concentrate on the scalar part of the propagator
(the spin indices can be added, but the scalar part controls the essential physics,
as in the canonical case of Fermi liquids \cite{Polchinski:1992ed}).
We first discuss equation (3) which describes the metaparticle propagator in the case when the
dual momenta are null, the propagator taking a non-relativistic appearance.
Imagine a Fermi gas of metaparticles, characterized by the Fermi energy $E_F$, and imagine that interactions between
metaparticles are turned on adiabatically \cite{anderson1}, and  that the resulting physics
is described by renormalizing towards the Fermi surface  \cite{Polchinski:1992ed}. Then if the term $p_{\perp}^2$
in equation (3) is replaced by energy $E$ measured from the Fermi surface, so that
%\be
$p_{\perp}^2 \to E$
%\ee
and the $p_{-}$ is set to be of the same order as perpendicular momentum 
%\be
$p_{-} \sim p_{\perp} \to \sqrt{E}$,
%\ee
which is natural close to the Fermi surface in the
limit of ordinary metals, we obtain that the $\mu$ term in Eq. (3) is
multiplied by $\sqrt{E}$, leading to an effective non-analytic behavior.
By considering Eq. (3) we observe that the effective quasi-metaparticle propagator interpolates between the
SYK model of strange metals and the usual Fermi liquid behavior.
In the SYK model of strange metals we have the two point function with 
$1/(\sqrt{E - E_F} -i\epsilon)$ dependence \cite{Sachdev:1992fk}
(see also, \cite{Patel:2018zpy}), which is to
be contrasted to the Fermi liquid behavior with dependence $1/(E -E_F - i\epsilon)$. 
Thus the Green's function that follows from eq (3) reads
\be
G_{MP} (E) \sim \frac{1}{(\mu_R \sqrt{E-E_F} + E - E_F - i \epsilon) }
\ee
after reabsorbing ${\cal E}$ into $\mu$ so that the rescaled $\mu_R$ becomes
%\be
$\mu \to \mu_R = \frac{\mu}{{\cal E}}$.
%\ee
Therefore, in the $\mu_R=0$
limit we obtain the usual quasiparticle Fermi liquid result. In the opposite limit (when the 
$\mu_R$ term dominates)
we obtain a strange metal, with a non-Fermi liquid behavior described by a particular SYK result.
In general we have the quasi-metaparticle Green's function
that interpolates between the Fermi liquid metal and strange metal (non-Fermi-liquid) limits.
(For other non-Fermi liquid behaviors consult, for example,  \cite{anderson2, Hartnoll:2009ns, Sur:2013wka}.)

Let us apply the ARPES probes to the quasi-metaparticle Green's functions.
The ARPES crucially probes the density of states \cite{varmareview}, which is related to the
imaginary part of Green's function. In the large $\mu_R$ limit 
this concept coincides with the SYK model \cite{Sachdev:1992fk} and the marginal Fermi 
liquid phenomenology \cite{varma, varmareview}. 
{Let us discuss whether metaparticles are consistent with the ARPES data collected to date
\cite{varmareview} (see also, \cite{georges}).} We follow Varma's insighful review \cite{varmareview},
specifically, section $B$ of that review, in which single-particle scattering rates by ARPES experiments are discussed. 
The imaginary part of the single particle self-energy
$Im \Sigma (E)$ is found to be $Im \Sigma (E) = a + b E$, where $b=0.7 \pm 0.1 $
is extracted from the data of \cite{arpes}.
In our case that means that  (for small $E$)
$\mu_R \sqrt{E-E_F} \sim \pm i \mu_R \sqrt{E_F} (1 - \frac{E}{2E_F})$, and thus
$\frac{\mu_R}{2 \sqrt{E_F}} =b = 0.7 \pm 0.1$, and 
$\mu_R \sim (1.4 \pm 0.2) \sqrt{E_F}$.
Thus the quasi-metaparticle concept is consistent with ARPES data and it captures
the non-Fermi-liquid behavior described by the imaginary part of the single particle self-energy,
due to non-trivial interactions in the standard description \cite{varmareview}.

Note that the usual Fermi-liquid result $1/(E -E_F - i\epsilon)$ leads to the textbook $T^2$ dependence of the resistance \cite{abrikosov}.
In contrast, the SYK Green's function $1/(\sqrt{E -E_F} -i\epsilon)$ leads to the observed linear $T$ dependence of the
resistance that is characteristic of strange metals \cite{varmareview}.
From the phase space argument it seems natural that if $1/E$ leads to $T^2$
then $1/{\sqrt{E}}$ should lead to the linear $T$ dependence. This phase space argument is a bit general, and from a more detailed point of view, we
envision a momentum dissipation mechanism that is responsible for the linear $T$ dependence that crucially involves
the dual momenta in the null gauge. However such a detailed discussion is beyond the scope of this letter.

What is the physical meaning of the parameter $\mu_R$?
We can answer this question by consulting the original 
paper by Sachdev and Ye \cite{Sachdev:1992fk}.
In particular the equation 8 of \cite{Sachdev:1992fk}
yields a Green's function which is precisely of the type discussed above
when the $\mu_R$ is large.
The relevant SYK result reads
%\be
$G_B (z) \sim \frac{i \Lambda e^{-i \theta}}{\sqrt{z}}$,
%\ee
where the parameter $\Lambda$ is explicitly stated as follows
%\be
$\Lambda = [ \frac{\pi}{J^2} \sin{2 \theta}]^{1/4}$,
%\ee
and hence we have that
%\be
$\mu_R \sim \Lambda^{-1}$,
%\ee
where the SY $\Lambda$ is given by equation 10 of \cite{Sachdev:1992fk} in terms of
the defining SY parameters.
This means that $\mu_R$ is essentially the strength of the disorder in the coupling, in this example, of disordered Heisenberg magnets. 
It seems like this would also mean that metaparticles with null $\tilde{p}$ are Schwinger bosons in the large $N$ limit. 
For a more recent discussion, involving numerical simulations,
\cite{georges} provides a discussion of this inverse square root behavior of the
Green's function in the SYK model.
A question arises about comparing the results to limit $B$ or limit $C$ of 
\cite{Sachdev:1992fk}. There are 3 parameters: $J$ (disorder strength), $n_b$ (related to the total "spin"), and the $M$ from $SU(M)$. In both limits, $M$ is taken to infinity. Note that $n_b$ is kept fixed in limit $B$, but diverges with $M$ in limit $C$. Assuming metaparticles are relevant to both limits, then the value of $\mu_R$ will depend on which limit is considered. In limit $B$, $\mu_R$ will depend only on $J$ (and the emergent parameter $\theta$), while in limit $C$, it will depend on both $J$ and $n_b$. In the former case, large 
$\mu_R$ would then correspond to large $J$, while in the latter it is less clear because even if $J$ is not large, $n_b$ still is, which is apparently enough to make 
$\mu_R$ large in this case.
Also, what happens as $\mu_R$ goes to zero is very subtle. For both limits, it seems this would imply that $J$ goes to zero, which 
represents a non-interacting "spin" system. 
Note that the SYK parameters $M$ and/or $n_b$ do not explicitly enter into the metaparticle propagator, and thus $\mu_R$ captures only an effective physics implied by these
specific parameters of the SYK model.
Note also that it is not clear that the SYK model can lead to the
Fermi liquid (quasiparticle) picture in any limit.
The metaparticle concept is compelling because
the Fermi liquid (quasiparticle) limit is apparently 
possible for $\mu_R$=0.
In other words, one has an interpolation between two deeply different physical domains.
Perhaps a
layered structure of regular metals and strange metals might exhibit such an interpolating behavior.

The quasi-metaparticle hence answers the general question: 
If a Fermi surface is given, what is the low energy excitation near the Fermi surface as all momenta are scaled to the Fermi surface?
The answer, in general, would be - a quasi-metaparticle.  In a singular limit 
of $\mu_R=0$, one recovers  the usual, Landau’s, 
answer -  the quasiparticle.
But for a non-zero $\mu_R$ 
one does not obtain a quasiparticle at all.
And in the large $\mu_R$ limit one can map to an effective behavior of the SYK model in a particular limit. 
In principle the $\mu_R$ parameter should be independently measured 
for a given system with a Fermi surface, and given its value, 
as compared to the Fermi energy,
one would find either a Fermi liquid metal or a strange metal 
or an intermediate state.

{\it Comments on High-temperature Superconductivity:}
Given this new picture of a generic effective behavior around a Fermi surface
represented by quasi-metaparticles, 
which only in a singular limit 
$\mu_R=0$ correspond to the canonical quasi-particles, one can ask whether
this new concept is useful for understanding the high-temperature superconducting
phase.
Here we note the following:
the simple one loop diagram computed in
Polchinski's lectures on the effective theory of Landau’s Fermi liquids 
\cite{Polchinski:1992ed} (see also \cite{abrikosov})
describes the essential physics of Cooper pairs 
(assuming an attractive interaction between the quasiparticles).
In what follows we repeat this computation in the context of quasi-metaparticles.

Essentially, one needs to compute, instead of the usual
\be
V^2 \int dE' \frac{d^2 p}{(2\pi)^4} dl' G_P (E+E', l')G_P (E-E', l'),
\ee
where the particle Green's function $G_P (E, l) \sim 1/(E -v_F l - i\epsilon) $
and where $v_F \equiv \partial_p E$ is the Fermi surface velocity, after the customary
scaling towards the Fermi surface \cite{Polchinski:1992ed},
the following analogous quasi-metaparticle expression
\be
V_{MP}^2 \int dE' \frac{d^2 p}{(2\pi)^4} dl' G_{MP} (E+E', l')G_{MP} (E-E', l'),
\ee
where the metaparticle Green's function is in general
$G_{MP} (E, l) \sim 1/(\mu_R \sqrt{E-v_Fl} + E -v_F l - i\epsilon) $,
and in particular, where one concentrates on the large $\mu_R$ limit.
Note the dimensionless nature of the interaction $V$ and $V_{MP}$.

The usual metalic one loop expression for the running
of the interaction vertex $V$ that results from the above one loop integral
\cite{Polchinski:1992ed} reads (to the leading logarithmic dependence) as
%\be
$V(E) = V - V^2 N (\ln(E_0/E) + O(1)) + O(V^3)$,
%\ee
where following \cite{Polchinski:1992ed}, $N$ is the density of states at the Fermi surface
%\be
$N = \int  \frac{d^2 k}{(2\pi)^2} \frac{1}{v_F (k)}$.
%\ee
This implies the one loop renormalization group equation
%\be
$E \partial_E V(E) = N V^2 +O(V^3)$,
%\ee
and thus 
%\be
$V(E) = \frac{V}{1+ NV \ln(E_0/E)}$,
%\ee
which for attractive $V <0$ increases in the infrared and leads
to the formation of Cooper pairs and the gap equation of the BCS theory.

In the quasi-metaparticle case we can also extract the leading logarithmic behavior,
obtaining
\be
V_{MP}(E) = V_{MP} - V_{MP}^2 N_{MP} (\ln(E_0/E) + O(1)) + O(V_{MP}^3),
\ee
where now
%\be
$N_{MP} = \int  \frac{d^2 k}{{\mu_R}^2 (2\pi)^2} \frac{{E_F}}{v_F (k)}$,
%\ee
as well as the following one loop renormalization group equation 
\be
E \partial_E V_{MP}(E) = N_{MP} V_{MP}^2 +O(V_{MP}^3),
\ee
and thus 
\be
V_{MP}(E) = \frac{V_{MP}}{1+ N_{MP} V_{MP} \ln(E_0/E)}.
\ee
This again leads to a Cooper pair formation mechanism (and the gap equation) for an attractive $V_{MP}$.
In this calculation, we have neglected the fundamental length and
non-commutativity, to first approximation.
Here an explict dependence of $N_{MP}$ on $\mu_R$ should be acknowledged.
This can be tied to a high critical temperature $T_c$ in high-temperature superconductors that have strange metals as their normal phases, because of the increased density of states
due to a complicated Fermi surface.
In general, we have that \cite{abrikosov},
%\be
$\exp(\frac{1}{NV}) \sim const + \frac{E_C}{T_C}$,
%\ee
where $T_c$ is also directly related to the gap at the Fermi surface \cite{abrikosov}
(see also, \cite{Minic:2008an}).

Therefore, if one replaces the canonical Fermi liquid propagator 
with the metaparticle propagator, we obtain a Cooper-like pairing even when we have 
a square root of energy in the propagator.
This is readily seen by considering the terms 
that would
give a logarithmic behavior for the above integral - actually the square root
of energy in the propagator is the only way to get a logarithm (and thus Cooper pairing) apart from the usual Fermi liquid case. 
This would imply that the metaparticles can be paired into Cooper pairs and lead to a superconducting state. So this particular metaparticle propagator that leads to a strange metal normal phase in a certain limit, can also lead to a superconducting phase. 
In an intuitive picture then, the normal phase is described in terms of
a non-Fermi liquid with metaparticle (little strings) excitations, which can form Cooper pairs, and thus form a superconducting phase.
Note that $\mu_R$ enters into the expression for the logarithm and thus the
high transition temperature is dependent on this parameter as well.

In this computation the fundamental length (stringy $\alpha'$) is set to zero,
but the other metaparticle parameters are taken to be non-zero.
Thus effectively the space-time and its dual commute in this simple
computation.
In general the interaction vertex would depend on the fundamental length and in that
case one has to deal with the explicit non-commutative field theoretic description
in which the logarithm would depend both on the UV and IR cutoffs  \cite{Freidel:2017nhg, Freidel:2017wst}.
In general one has a self-dual fixed point in this case, in order to properly define
the continuum limit, and take into account, properly, the UV/IR mixing
characteristic of such one loop computations in the non-commutative field theory
context \cite{Freidel:2017xsi, Douglas:2001ba}.
{One physical image of the metaparticle is that of two correlated particles - a particle
and its dual  \cite{Freidel:2018apz}. This might imply that the Cooper pair of metaparticles is naturally a
d-wave superconductor \cite{varmareview}.}

{\it Other remarks:}
Given the dimensionless nature of $\frac{\mu_R}{ \sqrt{E}}$, we can compute its one-loop correction, following
the above computation of the beta function for the interaction vertex $V_{MP}$. The one loop correction is
logarithmic in energy, and thus we have the following correction, $E \ln(E_0/E)$,
to the bare $\mu_R \sqrt{E}$ in the propagator. This functional form,  $E \ln(E_0/E)$, is encountered in the marginal Fermi liguid phenomenology \cite{varmareview},
as well as in the SYK model \cite{Sachdev:1992fk}. Thus, the dressed quasi-metaparticle propagator is also able to reproduce this
important feature of the SYK model and the marginal Fermi liquid phenomenology.

We further comment on the space-like and time-like case of the metaparticle propagator.
(Note 
the space-like case and its IR physics has been addressed in 
%\cite{Freidel:2018apz} 
\cite{Freidel:2021wpl}
in the context of dark matter and relativistic cosmology.)
In particular, suppose we take $\tilde p$ to be space-like, $\tp^\mu={\cal P}\tilde{n}^\mu$, where $\tilde{n}^2=1$, and we  can parameterize $p$ as $p^\mu=(p\cdot \tilde{n}) \tilde{n}^\mu+p^\mu_\perp$ with $\tilde{n}\cdot p_\perp=0$ and similarly for $x$. This means in particular that $p_\perp$ can be time-like. The propagator (stripped of the $\delta$-function) then reads \cite{Freidel:2018apz}
\beqn\label{metaPropSpaceliketp}
G(x,{\cal P},\tilde{n})
\sim
%\frac{e^{i\frac{\mu}{\cal E} x\cdot \tilde{n}}}{{\cal E}} 
\int \frac{d^{d-1}p_\perp}{|{\cal P}|} \frac{e^{i\left(\frac{\mu}{\cal P}\tilde{n}+p_\perp\right)\cdot x }
}{p_\perp^2+({\cal P}-\mu/{\cal P})^2 + m^2+2\mu-i\varepsilon}.
\eeqn
We see that the effect of the ${\cal D}$-constraint is to effectively fix the component of the momentum parallel to $\tp$.
Somewhat more difficult to interpret is the case of time-like or null $\tp$.
Indeed, suppose that $\tp$ is
time-like.
We write $\tp^\mu={\cal E}\tilde{n}^\mu$, where $\tilde{n}^2=-1$ and $\cE$ will be referred to as the dual energy.
We parameterize $p$ as $p^\mu=- (p\cdot \tilde{n}) \tilde{n}^\mu+p^\mu_\perp$ where $p_\perp$ is space-like and satisfies $\tilde{n}\cdot p_\perp=0$, and similarly for $x$. The propagator  reads \cite{Freidel:2018apz}
\beqn
G(x,{\cal E},\tilde{n})
\sim
%\frac{e^{i\frac{\mu}{\cal E} x\cdot \tilde{n}}}{{\cal E}} 
\int \frac{d^{d-1}p_\perp}{|{\cal E}|} \frac{e^{i\left(-\frac{\mu}{\cal E}\tilde{n}+p_\perp\right)\cdot x }
}{-\left({\cal E}-\mu/{\cal E}\right)^2  +p_\perp^2 +m^2-2\mu -i\varepsilon}.
\eeqn
The fascinating question here is: Do these other frames (spacelike and timelike)
for the metaparticle  correspond to
some new phases of matter in which such quasi-metaparticle excitations are important?

{\it Conclusion:} 
In this letter we introduced the concept of {quasi-metaparticles} based on the theory of metaparticles, the zero modes
of the metastring  \cite{Freidel:2017nhg, Freidel:2017wst}. We applied the concept of quasi-metaparticles to the problem of non-Fermi liquids and
the properties of strange metals. In particular, we pointed out that the quasi-metaparticle Green's function 
interpolates between the quasi-particle result and the result found in the context of
SYK model, which presents an exactly solvable model without quasiparticles. The linear dependence of
resistivity with temperature is reproduced in the SYK limit.
We note that metaparticles, understood as zero modes of the metastring  \cite{Freidel:2017nhg, Freidel:2017wst, Freidel:2018apz},
have extra symmetries 
(emergent from the many body viewpoint, see \cite{senthil})
such as remnants of scaling from the string world-sheet and T-duality \cite{Freidel:2018apz}. Also the full geometry of the metaparticle is
the Born geometry, found as well in the foundations of quantum theory  \cite{Freidel:2016pls}. 
Thus metaparticles and quasi-metaparticles should be quite robust in generic many body quantum systems.
As argued in this letter, the quasi-metaparticles can naturally describe the normal phase of high-temperature
superconductions and the Cooper mechanism is also possible for these new non-local
excitations.
An interesting question is whether the concept of quasi-metaparticles has a natural holographic dual,
given the holographic interpretation of the SYK model \cite{Sachdev:2010um} 
Another question concerns the construction of an explicit Hamiltonian for quasi-metaparticles that leads to
equation (4).
We hope to address these and other related questions in our future work.

{\it Acknowledgments:}
We thank Sean Hartnoll and Sung-Sik Lee for comments.
E.B. acknowledges support from the National Science
Foundation, grant no. DMR-1847078.
{\small DM} thanks the Julian Schwinger Foundation 
and the U.S. Department of Energy
(contract DE-FG02-13ER41917) for support.
{\small DM} thanks Perimeter Institute for
hospitality and he thanks Laurent Freidel, Rob Leigh and Jerzy Kowalski-Glikman for many inspiring discussions and
Philip Phillips for important suggestions during the recent remote workshop on strange metals.

%\bibliographystyle{uiuchept}
%\bibliography{metastring}
%\providecommand{\href}[2]{#2}\begingroup\raggedright

%\endgroup

\end{document}